# Wake Structure of Wind Turbines in Yaw under Uniform Inflow Conditions

Michael F. Howland[1], Juliaan Bossuyt[1,2], Luis A. Martínez-Tossas[1]
Johan Meyers[2], and Charles Meneveau[1]

28th January 2016

[1]Johns Hopkins University, Baltimore, MD 21218
[2]KU Leuven, Leuven, B3001, Belgium
Email: mike.howland13@gmail.com

**Abstract**

Reducing wake losses in wind farms by deflecting the wakes through turbine yawing has been shown to be a feasible wind farm controls approach. Nonetheless, the effectiveness of yawing depends not only on the degree of wake deflection but also on the resulting shape of the wake. In this work, the deflection and morphology of wakes behind a wind turbine operating in yawed conditions are studied using wind tunnel experiments of a wind turbine modeled as a porous disk in a uniform inflow. First, by measuring velocity distributions at various downstream positions and comparing with prior studies, we confirm that the non-rotating wind turbine model in yaw generates realistic wake deflections. Second, we characterize the wake shape and make first observations of what is termed a *curled wake*, displaying significant spanwise asymmetry. The wake curling observed in the experiments is also reproduced qualitatively in large eddy simulations using both actuator disk and actuator line models. When a wind turbine is yawed for the benefit of downstream turbines, the asymmetric shape of the wake must be taken into account since it affects how much of it intersects the downstream turbines.

## 1 Introduction

Considering the U.S. Department of Energy 20% Wind by 2030 plan [1] and similar goals elsewhere in the world [2], the efficiency and control of wind turbines placed in large wind farms has become an important area of study. Inevitably, significant power degradation occurs due to strong wake interactions between respective turbines downstream of each other [3–6]. Better understanding of these interactions is needed for improved designs of large, base load supplying wind farms. Currently, wind farms operate on the principle of maximum power point extraction, which entails each turbine to operate individually in an effort to maximize its own power at any time [7]. This operation can be considered similar to the control of a single, independent wind turbine that is not in a wind farm array. However, since such control strategies do not take wake interactions, and spatial or temporal correlations explicitly into account, they are most likely not the most effective strategy for an entire wind farm [8, 9]. Recently, there has been a push towards the optimization in the control of power generated by an entire large wind farm, as opposed to operating each turbine in a maximum power point tracking manner [10, 11]. In this vane, the wake deflection by operating wind turbines in yaw has been shown to be an attractive option to control wake deflection and power output [10, 12–16], and has generated significant interest recently [9, 17, 18].

Nominally, turbines are operated with the rotor perpendicular to the flow, with tip speed ratio and pitch near optimal values, which are dependent on the turbine and the desired power output. In an effort to reduce the power losses for downstream wind turbines that reside in the wake of an upstream one, there have been experimental studies which have considered altering yaw angle, tip speed ratio, and blade pitch [14, 17, 18].

Ref. [14] used two aligned turbines in a wind tunnel and tested varying the rotor yaw angle, tip speed ratio, and the blade pitch of the upstream wind turbine only. This study showed that varying the yaw angle of the wind turbine was of comparable benefit to increasing the streamwise spacing between turbines, with an optimal power output occurring at 30°. Refs. [17, 18] studied the effects of controlling yaw angle, tip speed ratio, and the blade pitch of the upstream turbine for scaled model wind turbines, with results also revealing the benefits of yawing the upstream turbine. Further, yaw misalignment has been shown to reduce the steady-state blade loading variations by up to 70%, which has lead to the use of yawing to increase operational life [19]. Ref. [20] studied a rotating wind turbine model in replicated atmospheric boundary layer conditions to discover a deflection of approximately $0.6D$ in the far wake.

Refs. [9, 21–23] were computational studies of wake deflection using various yaw angles. Ref. [21] uses LES with an actuator disk model with turbulent inflow and shows that wake deflection can be reproduced in such simulations. They also propose a momentum-based model for the deflection which is compared to LES with reasonable validity in the far wake. Some experimental results are compared, but the authors cite a need for more experimental verification before a wake controller may be developed.

Ref. [9] studied wake deflection under various conditions using the SOWFA Large Eddy Simulation (LES) code and using the NREL 5 MW turbine model [24]. When the yaw angle $\gamma$ was $\gamma = 30°$, the study found the maximum wake deflection to reach about 0.5D in the far wake, where $D$ is the rotor diameter. Ref. [22] studied the near wake structure of a wind turbine under uniform inflow using Reynolds Averaged Navier-Stokes flow modeling and the results displayed some strong asymmetries in the near field (up to $2D$ downstream). Furthermore, employing an actuator disk model for the turbine under uniform shear, Ref. [23] found wakes deflected up to $0.7D$ when $\gamma = 30°$. Further LES studies of several yawed turbines have been carried out in Ref. [25], and they compared the wake deflection with the theoretical model of Ref. [21], which characterizes the skew angle behind a yawed turbine.

Most of the studies considered only 2D wake deflection in horizontal planes, generally at hub height. However, the wakes of wind turbines have been shown to exhibit asymmetric properties in yaw, as pointed out in Ref. [26]. The spanwise forcing imposed by a wind turbine operating in yaw has been shown to be significant. Additionally, Ref. [26] has noted the importance of free stream turbulence on the structure of the 3D wake, which influences the high energy mixing downstream.

In general, prior studies have shown that yawing turbines has power reduction for the yawed turbine (following $cos^3(\gamma)$), but can yield noticeable power increases for nominally aligned downstream wind turbines as a result of the deflected wake. Even when wind turbines operate nominally in non-yawed conditions, in practice there always is some yaw misalignment due to the imperfections of the yaw control for aligning the turbine with the incoming wind. In fact it has been shown with LIDAR measurements that wind turbines typically operate from 4° to 10° in yaw when the turbine attempts to track the flow to operate with 0° yaw [27]. Therefore, understanding of the dynamics and implications of a wind turbine operating in yaw are important to the design and control of wind farms even if traditional yaw alignment controllers are used.

The objectives of this study are to examine the use of drag disk type model wind turbines for the use in wake deflection experiments. For experimental studies of large wind farms, it is often desirable or necessary to use non-rotating porous disk models, in order to accommodate a large number of small model turbines that may be installed within the physical constraints of typical wind tunnels [28]. As such, the mechanism of wake deflection when using a porous (or actuator) disk model must be established in order to enable further studies. To our knowledge, there has not been an experimental study of porous disk model turbines in yaw to study wake deflection. A wind tunnel experiment, described in §2, is performed and results are presented in §3, where the center of wake is defined and then determined from the data and compared with prior studies. Also, streamwise and spanwise mean velocity distributions are mapped out to characterize the shape of the wake at various downstream cross-sections with particular attention to the shape of the resulting wake, shown in §4. Traditional wake models assume a symmetric, circular shape but as will be shown, significant asymmetries develop in yawed wakes. In order to provide further evidence of the particular wake morphology determined experimentally, we perform large eddy simulations using both actuator disk and actuator line methods and confirm, qualitatively, the observed wake shapes. Large eddy simulations are presented in §5. Conclusions are presented in §6.

## 2 Experimental Setup

Experiments are performed in the Corrsin Wind Tunnel at the Johns Hopkins University. It is a closed loop, two-story facility, with a primary contraction-ratio of 25:1 and a secondary contraction of 1.27:1. The test section is 10 m long with a cross section of 1 m by 1.3 m. The experiments are performed in laminar, uniform inflow, with free-stream velocity in the test section of $U_\infty = 12\ m/s$. The free stream turbulence level is less than 0.12%. To ensure uniform inflow, the wind turbine model is placed far downstream of the contraction and in the center of the cross section, far from any walls (the boundary layer thicknesses at the measurement location are below 8 cm). The single turbine is mounted on a slender cylinder which is connected to a stepper motor with a step size of 0.1125° allowing precise control of the yaw angle. Overall, we estimate the systematic yaw uncertainty to be ±0.5° due to uncertainties in turbine placement within the experimental domain. The $x$, $y$, and $z$ coordinate directions are streamwise, spanwise, and height respectively and are shown in Fig. 1.

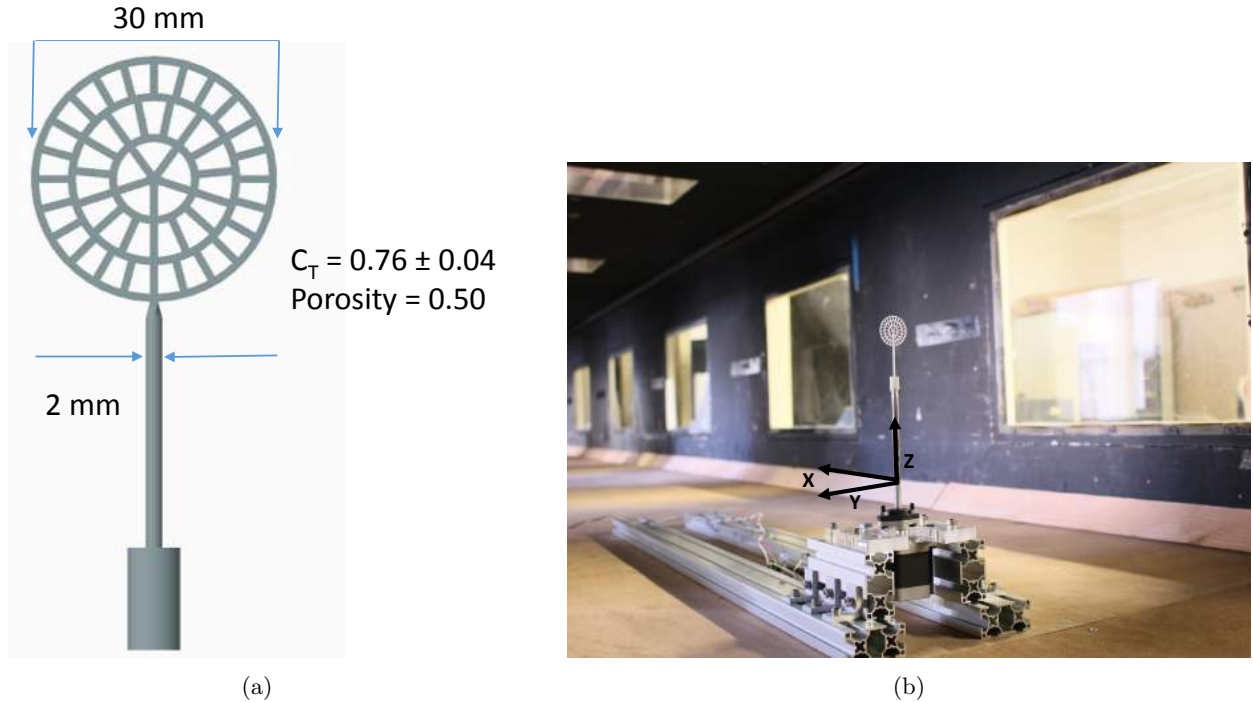

(a) (b)

Figure 1: Schematic of 3D printed drag disk model turbine (a), and photograph of the model turbine and yaw control stepper motor mounted in the JHU Corrsin Wind Tunnel (b).

Experiments use a porous disk wind turbine model which was designed to match the far wake properties of a full scale wind turbine [28]. Fig. 1 shows a schematic and a photograph of the porous disk and the setup in the wind tunnel. The diameter of the model turbine is 3 cm, i.e. a scale ratio of about $4 \times 10^3$ compared to a large-scale $D = 120m$ utility wind turbine. Such a scale ratio is needed here to fit 100 turbines inside the test section. It would be very challenging to build rotating model turbines of such small diameters that would still produce the correct thrust and induction coefficients and correct turbine control. These parameters mainly determine the overall properties of the wake. The turbine has been designed to match a desired thrust coefficient of $C_T = 0.75$ and is manufactured using 3D printing. Its properties have been carefully documented in Ref. [28] for the case of non-yawed conditions, showing excellent agreement with the desired thrust coefficient (measured using strain-gages) and canonical wake defect velocity profiles that agree very well with those of rotating wind turbines at streamwise distances beyond 3D.

Measurements are performed using hot wire anemometry and a pitot-static tube. The hot-wire meas-urements were made with with an X-wire probe made in-house as described in Ref. [29]. The probe is mounted on a three-axis traverse system with spatial location accuracy of ± 0.1 mm. Signals are acquired at a sampling rate of 10 kHz, with a low pass filter (Nyquist) of 5 kHz, capturing both the mean velocity and the variance of the velocity signal accurately. Signals are acquired at each measurement location for 26 seconds to ensure converged mean and second-order flow statistics. The X-wire is oriented such that the $u$ and $v$ components (streamwise and spanwise components, respectively) of the velocity are measured. In order to compensate for the temperature drift of the hot-wire probe measurement system, the data is recalibrated to $U_\infty$ when the probe is in the free stream, with subsequent measurements adjusted using linear interpolation, as done in Ref. [30]. Measurement locations along YZ and XY planes are shown in Fig. 2. XY planes were taken at hub height in order to characterize the 2D wake deflection. The YZ planes were taken at $x/D = 5, 8$ for the hot-wire probe in order to show the development of the wake structure in the far wake. Typical turbine placement is 5D - 8D, so the wake deflection and structure between these locations is important.

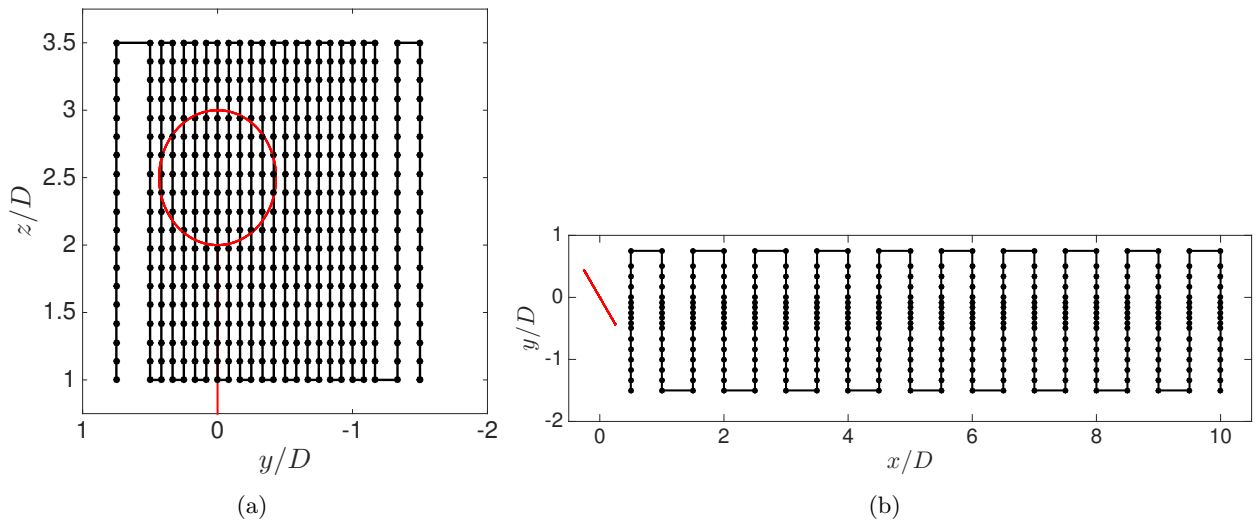

Figure 2: (a) Distribution of measurement points for the YZ plane experiments for the yawed turbine and (b) measurement points for the XY plane experiments. The YZ plane is viewed from the negative x direction and XY plane is viewed from the positive z direction. Red ellipse in (a) and inclined plane in (b) represents the corresponding two views of the yawed turbine.

The pitot measurements were carried out with a Pitot static tube with an outside diameter of 2 mm. The pressure was measured with an 220CD Baratron General Purpose Differential Capacitance Manometer with measurement uncertainty of ±0.15%, leading to an error of ±2 Pa. The output voltage was measured with an Omega Instrunet i555. Together, this setup results in an overall velocity measurement uncertainty of ±0.2 m/s in the case of 7 m/s laminar flow, the lowest velocity measured with the pitot setup in the wake of the turbine. This gives a maximum pitot velocity uncertainty of 3%. Added pressure effects due to turbulence will lead to a measurement offset in the wake of the turbine [31]. The pitot-static tube was used for an XY plane at hub height and YZ planes at $x/D = 0.5, 1.5, 2, 4, 5, 6, 7, 10$, and 12. The pitot probe is used for the characterization of the center of the wake, but will not be used for detailed velocity measurements. Pitot probes were chosen for wake deflection characterization since hot-wire measurements require a significantly more elaborate calibration process and have a higher sensitivity to temperature drift during long duration measurements [32]. In the wake of the turbine, the turbulence intensity is not uniform, which may alter the uncertainty of the pitot tube during the experiment. However, as further shown below, reasonable agreement between pitot and hot-wire probe was observed for the wake deflection characterization.

# 3 Center of Wake Deflection

With the turbine in yawed conditions, the wake is no longer symmetric in the spanwise dimension. Further, when the wind turbine tower is included, the wake is not symmetric in height either. As a result, it becomes necessary to characterize the center of asymmetric wakes in order to compare different yaw angles and control methods. Several methods have been proposed before, such as fitting a Gaussian shape [9, 21] or using the "center of mass" of the velocity defect [25, 33]. Additionally, Ref. [34] has proposed using particles to track the center of wake for turbines in yaw, yet this study only considers particles deflection in a horizontal slice, not the 3D wake effects. A 3D wake center also considers the tower wake. Since the wake shapes will be found to differ significantly from Gaussians and exhibits 3D properties, here we use the "center of mass" approach method. The center of the wake is computed at every streamwise distance in the flow, according to the resolution of the domain. At each streamwise measurement location $x$, mean streamwise velocity data on a YZ plane is considered. The center of wake coordinates $y_c(x)$ and $z_c(x)$ are computed according to

$$y_c(x) = \frac{\iint y \, \Delta U(x,y,z) \, \mathrm{d}y\mathrm{d}z}{\iint \Delta U(x,y,z) \, \mathrm{d}y\mathrm{d}z}, \quad \text{and} \quad z_c(x) = \frac{\iint z \, \Delta U(x,y,z) \, \mathrm{d}y\mathrm{d}z}{\iint \Delta U(x,y,z) \, \mathrm{d}y\mathrm{d}z}, \tag{1}$$

where $\Delta U(x,y,z) = U_\infty - \bar{u}(x,y,z)$, $\bar{u}$ is the time averaged velocity and $U_\infty$ is, as before, the free stream velocity. The integration is performed over the available spatial data.

To obtain the center of wake from the XY-plane measurements at the many $x$ locations, we use 1D integration in the y-direction only and neglect the z-dependence of the wake

$$y'_c(x) = \frac{\int y \, \Delta U(x,y,z=0) \, \mathrm{d}y}{\int \Delta U(x,y,z=0) \, \mathrm{d}y}, \tag{2}$$

In Fig. 3, filled circles represents $y_c(x)$ from pitot data in successive YZ planes at the various $x/D$ distances downstream. The cross markers show the $y'_c(x)$ computed from pitot data from an XY plane measurement at hub height. The open circles represents $y'_c(x)$ for hot-wire probe results for which data was available in an XY plane measurement at hub height. All measurements were traversed with the system described in §2 and point maps shown in Fig. 2. The experimentally measured wake deflection downstream for the $\gamma = 30°$ yawing case is compared with results from literature. Specifically, in Fig. 3 we compare the center of wake computed from Eq. 2 with pitot and hot-wire measurements and the center of wake computed from Eq. 1 from pitot measurements with wind tunnel results from Ecole Polytechnique Fédérale de Lausanne (EPFL) [20] and with numerical simulations from National Renewable Energy Laboratory (NREL) [9] and Danish Technical University (DTU) [23]. The different conditions are summarized in Table 1. Estimating the experimental uncertainty associated with the pitot and hot-wire probe measurements is challenging. For the pitot probe, we choose the maximum measured deviation of $y'_c(x)$ for a case in which the deflection should be identically zero (the case of zero yaw). More details are provided below. The uncertainty estimated in this fashion is approximately $\pm 0.06D$. For the hot-wire data, we assume a 2% error in velocity [29, 32] and the traverse positioning error described in §2, yielding an estimated error in $y'_c(x)$ of about $\pm 0.02D$ through standard propagation of error in Eq. 2. The wake center is approximately consistent between pitot and hot-wire probes, as seen in the characterization of $y'_c(x)$ in Fig. 3. In §4, a quantitative analysis of wake structure and statistics is performed with the hot-wire probe.

As can be seen in Fig. 3, considering the expected uncertainties involved in the present experiments (as represented by the error bars) and the slight differences in setup and conditions of the published results, the agreement is good. Results seem to suggest that for $x/D > 8$, the center of wake deflection asymptotes to about $\sim 0.6D$ for the 30° yawing case. This result verifies that the drag disk wind turbine model produces realistic wake deflection for a given yawing angle and thus justifies its use in wind farm yaw studies for wake deflection and power optimization. Furthermore, since we observe the deflection for a non-rotating turbine model, results confirm the findings of [21, 23] which argue that the magnitude of wake deflection is predominantly a function of $C_T$ rather than rotational effects of the turbine's blades. Moreover, Ref. [21] proposes a simple model for the wake deflection as a function of $C_T$. Fig. 4 shows that the final deflection corresponds well with Jimenez's model with $C_T$ values as in Table 1. Finally, the differences in $y_c(x)$ and $y'_c(x)$ are well characterized by the pitot probe data in Fig. 3, and will be further explored in §4.

Table 1: Comparison of the turbine models [9, 21, 23, 28].

|  | Porous Disk | NREL | DTU | EPFL |
|---|---|---|---|---|
| $C_T$ | $0.76 \pm 0.04$ | 0.9 | 0.64 | 0.85 |
| Background Turbulence Intensity | 0.12% | 6.30% | 10% | 7.50% |
| Diameter | 0.03 m | 126 m | 80 m | 0.15 m |
| $U_\infty$ | 12 m/s | 8 m/s | - | 4.88 m/s |

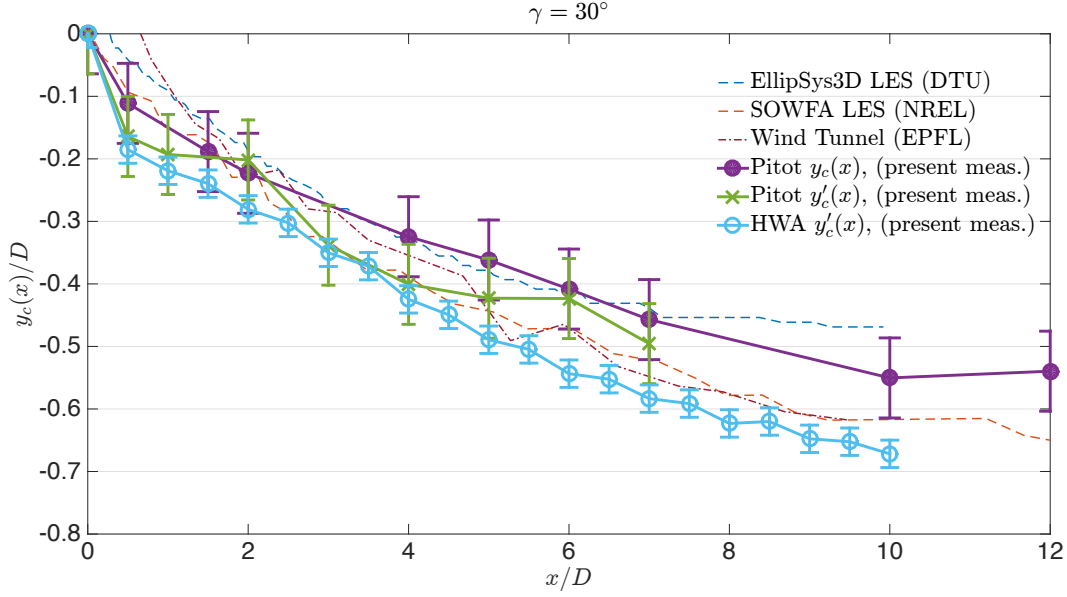

Figure 3: Comparison of the wake deflections for a yaw angle of $\gamma = 30°$ from Refs. [9, 21, 23] with present hot-wire measurements in the wake of a porous disc model in a wind tunnel. Present measures shown with $y_c(x)$ for Pitot probe data and $y'_c(x)$ for both Pitot and hot-wire probes. Error bars denote the experimental uncertainty, determined by combining the estimated uncertainties due to Pitot and hot-wire probes, traverse system, and yaw controller.

Further experiments for other yaw angles have been performed, for $\gamma = 0°, 5°, 10°, 20°$, and $30°$. For these cases, however, measurements of mean velocity were only performed in XY planes at hub height using a pitot-probe. Hence, no detailed velocity measurement are presented from the pitot probe results. However, as before, the resulting center of wake positions, $y'_c(x)$, being given by a ratio of integrated velocity distributions, are expected to be fairly insensitive to the inaccuracies of the pitot probe. The results are shown in Fig. 4.

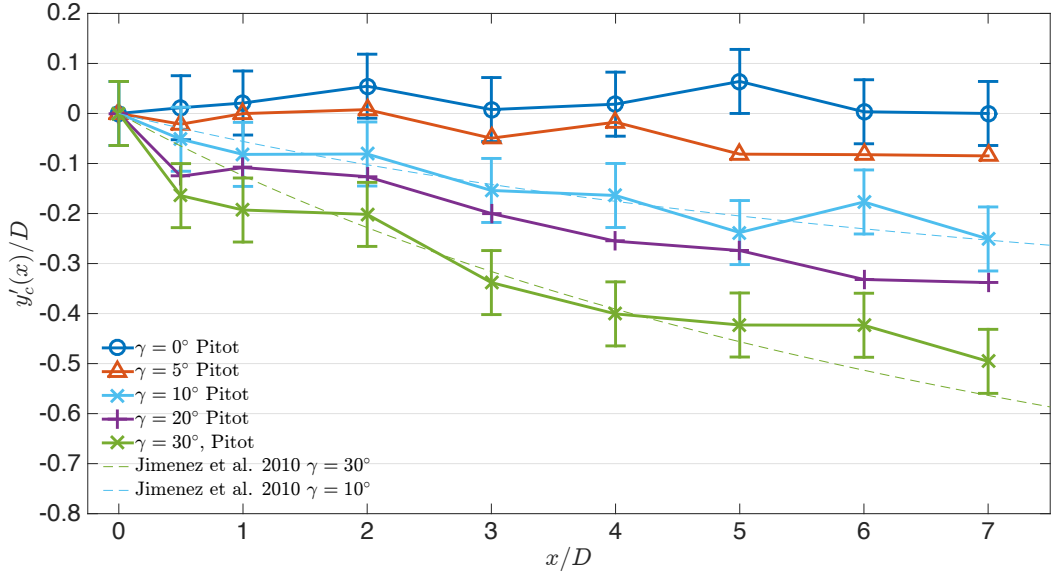

Figure 4: Comparison of the wake deflections ($y'_c(x)$) for $\gamma = 0°, 5°, 10°, 20°$, and $30°$ tracked in XY planes at hub height $z/D = 0$ with yawed wake deflection model described in Ref. [21] and given by Eq. 4. Errorbars shown for $\gamma = 0°, 10°$, and $30°$.

## 4 Wake Shape

To illustrate the 3D wake deflection of a wind turbine in yaw, we consider the shape of the wake. These results were acquired using the hot-wire experimental setup described in §2.

### 4.1 Streamwise velocity distributions

Fig. 5 shows the streamwise mean velocity distribution (normalized with free-stream velocity) on an XY plane at hub height of the wind turbine. It clearly reveals the wake deflection under yawed ($\gamma = 30°$) conditions. The wake center, $y'_c(x)$, computed previously is shown with full circles.

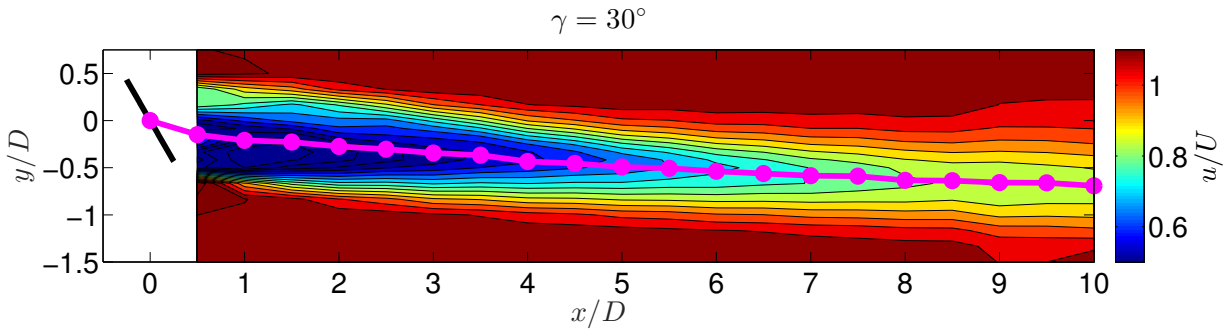

Figure 5: Time averaged streamwise velocity contour plot at hub height ($z/D = 0$), taken with a hot-wire probe. The mean velocity is normalized by free-stream velocity $U_\infty = 12$ m/s. The dark black line represents the yawed turbine. The XY center of wake $y'_c(x)$ is shown in filled magenta circles.

Next, we consider the shape of the wake in cross-stream YZ planes. Fig. 6 shows the mean velocity distribution at $x/D = 5$ and 8, normalized with free stream velocity ($U_\infty$). The wake has an asymmetric, curled shape. We will refer to this type of wake as the *curled wake*. As a result of its 3D shape, the momentum deficit region behind the yawed turbine is not fully deflected to the amount implied by the XY plane measurements, since the wake experiences maximum deflection at hub height. That is to say, $y'_c > y_c$. The wake is deflected considerably less towards the (negative) y-direction at the top and the bottom of the rotor area. Thus, care must be taken when characterizing the wake deflection based on $y'_c$ measured only at $z = 0$ since it may overestimate the overall deflection. Also, the wake of the tower is deflected in the opposite direction of the rotor wake. As will be seen below, the lateral deflection of the tower wake is a result of the spanwise mean velocity which below (and above) the rotor area points towards the positive y-direction.

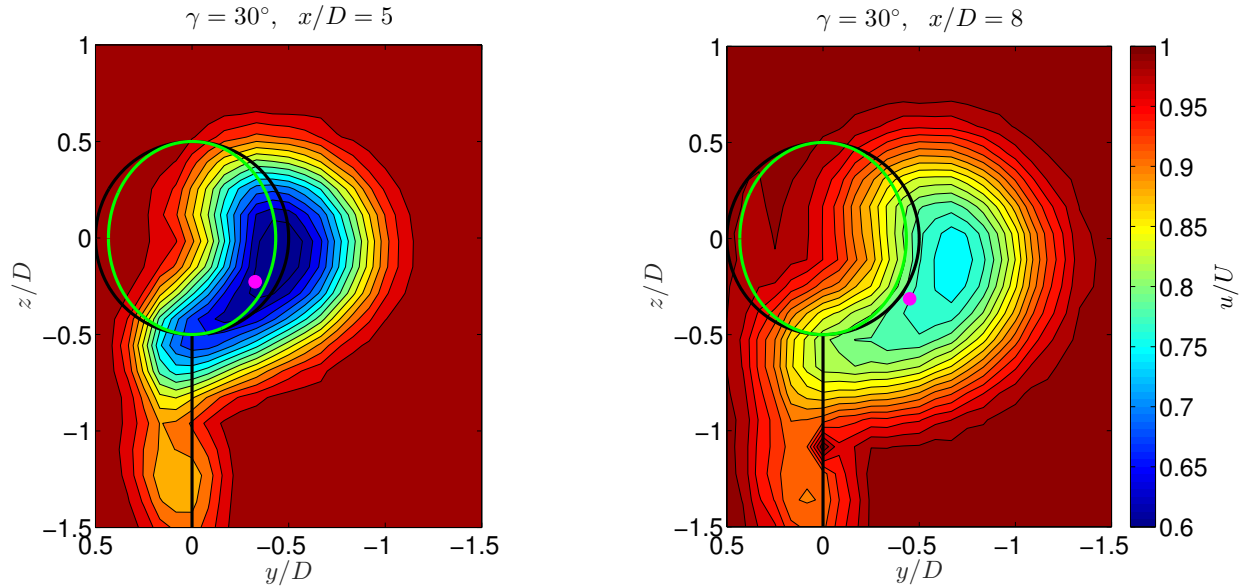

Figure 6: Time averaged streamwise velocity contour plot at $x/D = 5, 8$ downstream, taken with a hot-wire probe. The mean velocity is normalized by $U_\infty = 12$ m/s. The disk area projected on the YZ plane is shown in green. $y_c(x)$ is shown in magenta.

## 4.2 Spanwise Velocity

Fig. 7 shows the spanwise $v/U$ mean velocity distribution on the XY plane at hub height ($z = 0$). The velocity contours show the development of the strong spanwise velocity that deflects the wake of the turbine in yaw. The magnitude of the spanwise velocity near the centerline is relatively constant, about $(0.10 - 0.15)U_\infty$, until approximately 5D, and then slowly decreases.

The spanwise velocity contour plots on YZ planes shown in Fig. 8 for $x/D = 5$ and $x/D = 8$ suggest the mechanism for the development of the curled wake. In the center of the wake of the yawed turbine, there is spanwise velocity consistent with the sideways thrust applied by the yawed turbine. The center spanwise (negative) $v$-velocity transports the initial streamwise velocity defect towards the (negative) $y$-direction, the direction of the overall wake deflection. However, the degree of such "transport" is proportional to the $v$-velocity magnitude which decreases away from $z = 0$, thus leading to the curled shape of the wake. Interestingly, above the rotor area, at $|z|/D > 0.5$, the $v$-velocity is positive, i.e. in the opposite direction of the implied transverse thrust. Such flow direction suggests that the tilted disk is generating mean streamwise vorticity at its top and bottom edges. The positive $v$-velocity regions on the top and the bottom of the rotor area transport the wake velocity defect in the opposite (positive $y$) direction, thus further enhancing the wake curling.

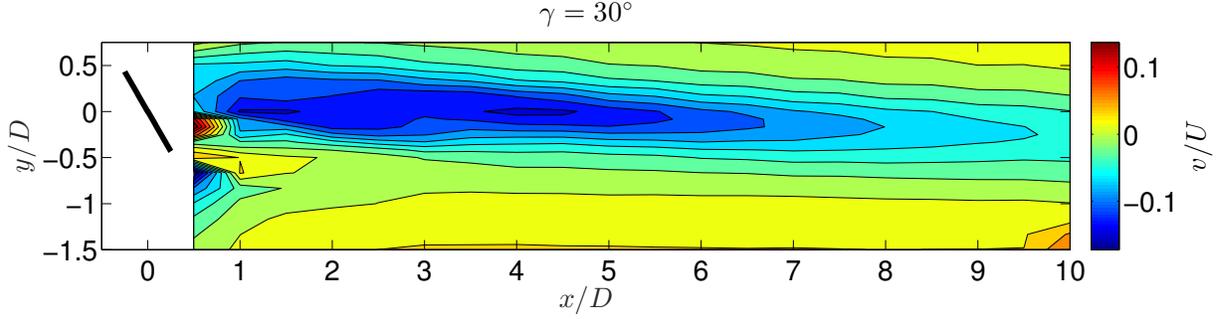

Figure 7: Contour plot of time averaged spanwise velocity at $z/D = 0$ (hub height), taken with a hot-wire probe. The dark black line represents the yawed turbine. The mean velocity is normalized by free-stream velocity $U_\infty = 12$ m/s.

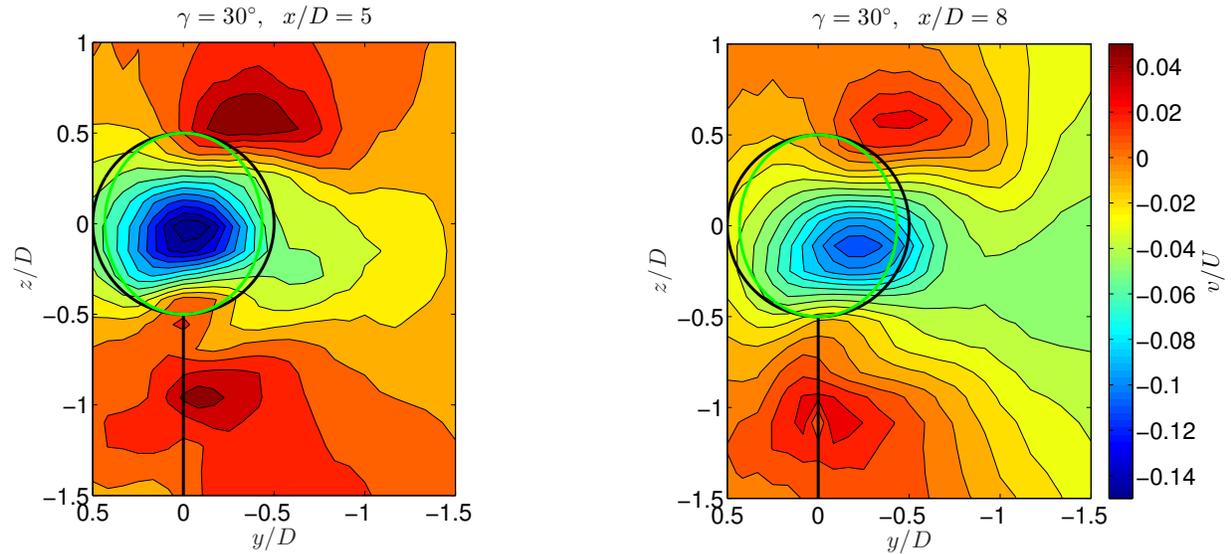

Figure 8: Time averaged spanwise velocity contour plot from hot-wire probe at $x/D = 5, 8$ downstream. The mean velocity is normalized by free-stream velocity $U_\infty = 12$ m/s. The disk area projected on the YZ plane is shown in green.

## 4.3 Turbulence Intensity

The turbulence intensity (in %) is defined as $TI_u = 100 \cdot \sqrt{\langle u'^2 \rangle}/U_\infty$ (i.e. normalized by the unique $U_\infty$) and is evaluated from the hot-wire data in YZ planes at $x = 5$ and $x/D = 8$. Resulting distributions are shown in Fig. 9. As can be seen, the turbulence intensity distribution in the wake of the yawed turbine also shows the development of the curled wake phenomenon. The maximum turbulence intensity is at hub height in the center of the deflected wake, while the overall shape is curled.

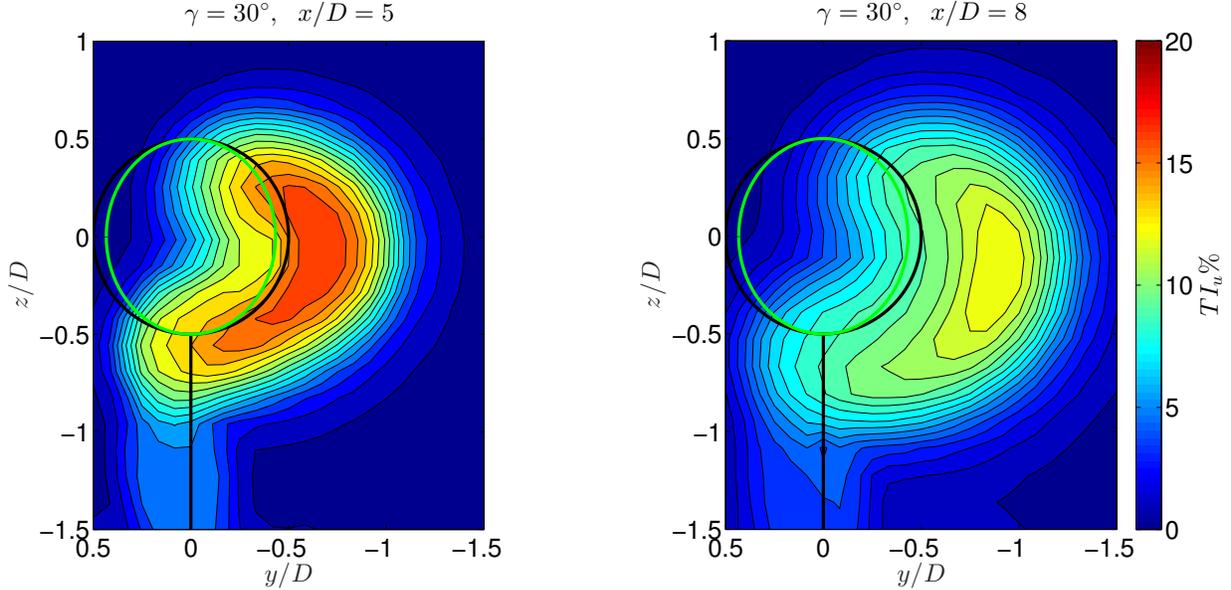

Figure 9: Streamwise turbulence intensity contour plots normalized by $U_\infty$ from hot-wire probe, at $x/D = 5, 8$ downstream. The disk area projected on the YZ plane is shown in green.

## 5 Large Eddy Simulations

In order to explore whether the experimentally measured curled wake phenomenon is also present in numerical simulations, we perform Large Eddy Simulations of a yawed turbine under uniform inflow. We use the JHU LES solver which has been used in a number of prior wind farm studies [35–39]. The code is a low dissipation psuedo-spectral solver. It is psuedo-spectral in two dimensions with the z-direction employing second-order centered finite differencing. The Scale Dependent Lagrangian subgrid scale model [40] is used. It has been compared to another LES code in Refs. [39, 41] and to single non-yawed turbine wake measurements in [42]. Present simulations are performed with a stress free boundary conditions on all side boundaries, with uniform, laminar inflow with $U_\infty = 12\ m/s$. $N_x$, $N_y$, and $N_z$ are 512, 128, and 256 respectively, with a domain size of 25D x 5D x 5D. The resolution in $z$ is twice the resolution in $x$ and $y$ to ensure consistent Reynolds stresses [39].

The turbine is modeled with the actuator disk model as described in Refs. [37, 43] and is placed at the center of the domain cross-section at $x = 5D$. A fringe region of 5% of the domain length was used to specify the inflow velocity in the context of the periodic $x$ direction boundary conditions of the code [37]. The yawed forces are computed using the unit normal vector in each dimension from the turbine, as also done in Ref. [21]. $\mathbf{f}'(\mathbf{x}, \mathbf{y}, \mathbf{z}) = \mathbf{f}(\mathbf{x}, \mathbf{y}, \mathbf{z}) \cdot \hat{\mathbf{n}}$, where $\mathbf{f}(\mathbf{x}, \mathbf{y}, \mathbf{z})$ is the non-yawed ADM force at each node within the turbine, $\hat{\mathbf{n}}$ is the unit normal in each direction $(\hat{i}, \hat{j}, \hat{k})$ from the turbine, and $\mathbf{f}'(\mathbf{x}, \mathbf{y}, \mathbf{z})$ is the resulting yawed force, i.e. $f'_i(x, y, z) = f_i(x, y, z) \cdot cos(\gamma)$.

The wind turbine tower was modeled as a drag based object which only forces in the streamwise direction. The small forcing in the spanwise dimension is neglected. The tower diameter $d_T/D = 1/15$ was used to specify the drag force (the same diameter ratio as in the wind tunnel experiments), with a drag coefficient of $C_D = 1$ in low Reynolds number flow [44]. A Gaussian kernel [39, 45] was used for both the wind turbine actuator disk and the tower, with a kernel width of $\epsilon = 2\Delta x = 0.0391D$.

Fig. 10 shows the mean streamwise velocity contours in the XY plane. The wake deflection is thus confirmed numerically from the LES results. As can be seen, however, the decay of defect velocity is significantly more gradual than for the experimental data. Moreover, we note from instantaneous plots (not shown) that the simulated wake becomes turbulent rather far downstream (not before $x/D \sim 7 - 8$).

This differs from the experimental results which show that the wake behind the porous disk consisting of a grid as shown in Fig. 1 is turbulent immediately. The turbulence occurs mostly at small scales initially comparable to the grid-spacing, thus helping to diffuse the wake more rapidly than in the simulations where the actuator disc applies a spatially uniform force. Several attempts were made to introduce random forcing at the rotor location to trigger earlier transition in the LES, but results were not satisfactory and dependent on the random forcing chosen. In most prior actuator disk model applications [21, 35, 37, 38] the inflow to the turbine was highly turbulent already and thus natural transition of a laminar wake was not an issue as it is for the present configuration. Since our main objectives are on qualitative features of the wake deflection and its shape rather than on a detailed quantitative code/experiment validation for a wind turbine in uniform inflow, further comparisons between the simulated and experimentally measured wake only refer to qualitative trends of wake deflection and wake curling.

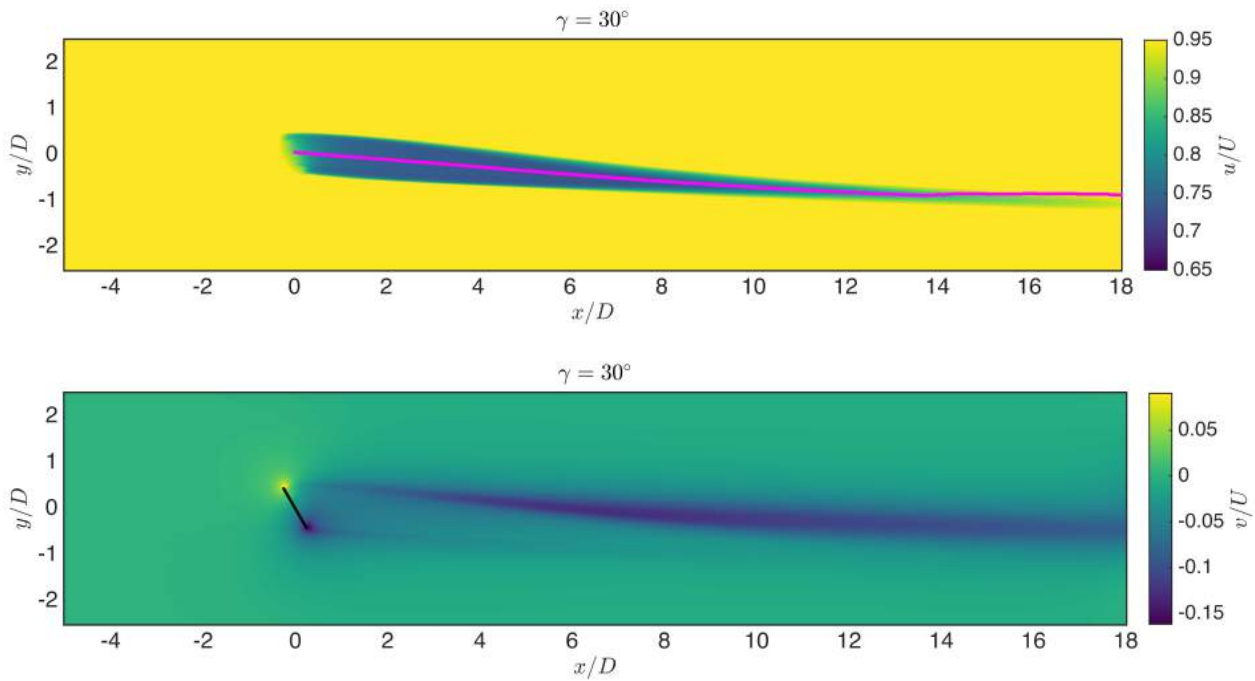

Figure 10: Time averaged mean velocity contours normalized by the free stream velocity on a XY plane at hub height $z/D = 0$ obtained from ADM LES. The XY center of wake $y'_c(x)$ is shown in magenta and (b) spanwise velocity at hub height.

The yawed turbine creates a set of counter-rotating vortices in the top and bottom of the rotor. This can be shown in Fig. 11 where the streamlines represent the velocity field components in the YZ plane. As the wake evolves downstream, these counter rotating vortices are responsible for shifting the wake from its center location. Further downstream, the wake obtains its *curled wake* shape. These vectors can only be seen in the LES simulations, where all the velocity components are computed. In the experimental measurements the $w$-velocity, the component in the $z$ direction, is not present.

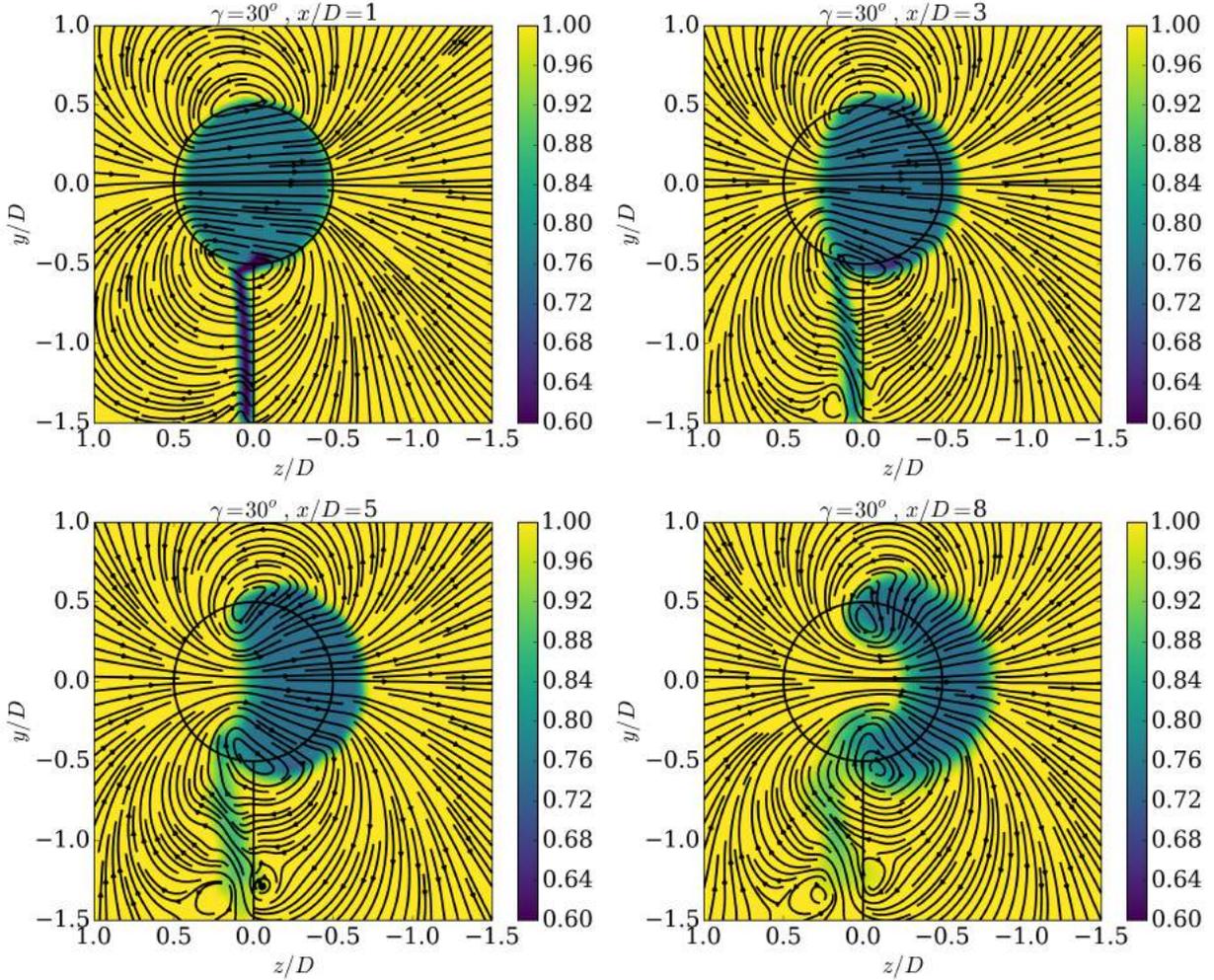

Figure 11: Streamwise velocity ($u/U_\infty$) contours from ADM LES with streamlines showing the vector components of the velocity field on the YZ plane for 1D, 3D, 5D and 8D.

Finally we also perform a simulation of a wind turbine using ALM implementation. For simplicity, we use the previously simulated case of a 5 MW NREL turbine as described in Ref. [24, 46]. It does not include a tower. For additional details about ALM, see Ref. [39]. The resulting streamwise velocity contours on cross-stream planes are shown in Fig. 12. As can be seen, the curled wake is also present in LES using the ALM. However, the wake shape exhibits some dissimilarities with the ADM and the porous disk, since the rotor rotation now also introduces top-down asymmetry into the flow. Nevertheless, the center of the wake deflection is similar to that obtained from the LES using ADM, and the curling is also observed.

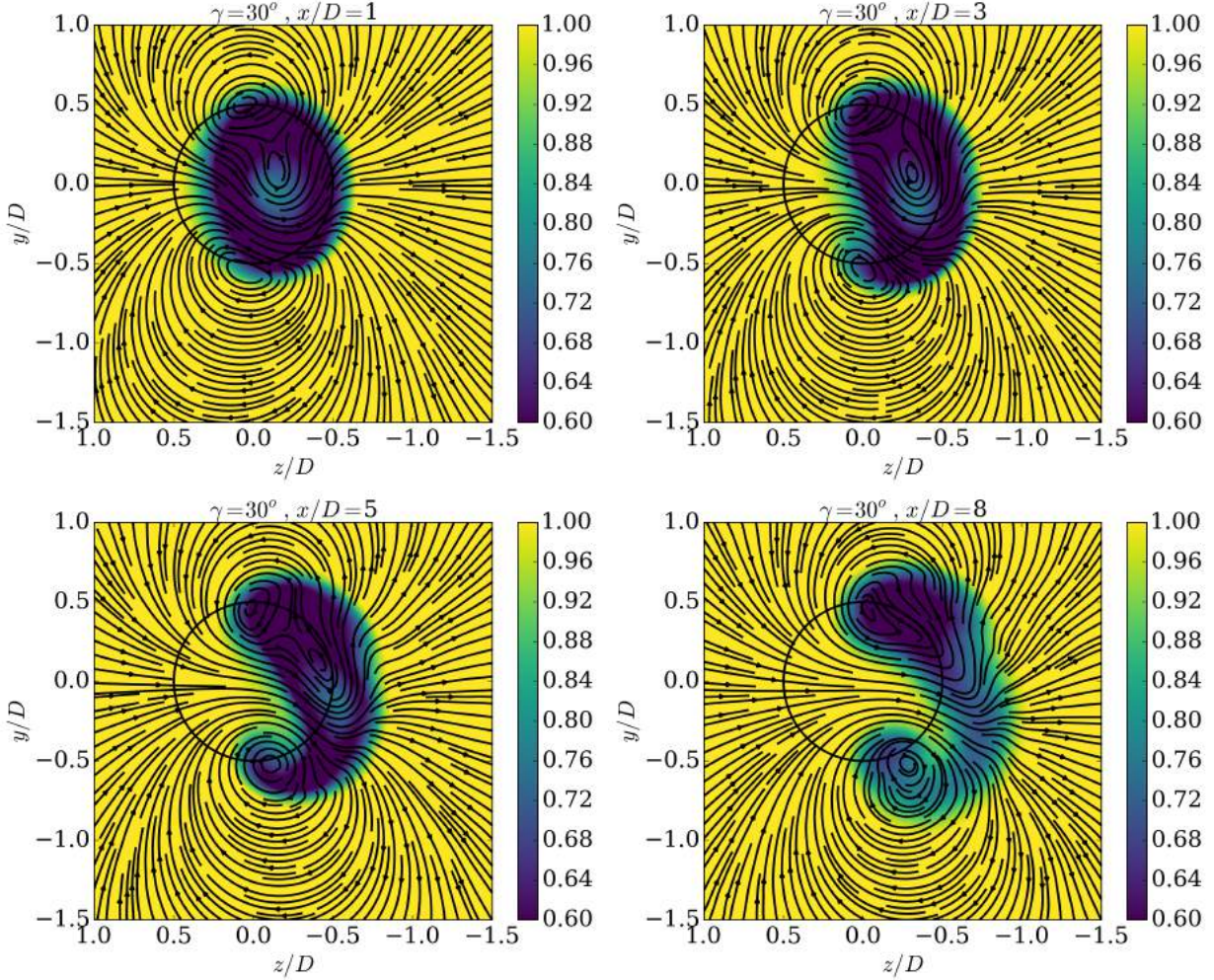

Figure 12: Streamwise velocity ($u/U_\infty$) contours from ALM LES with streamlines showing the vector components of the velocity field on the YZ plane for 1D, 3D, 5D and 8D.

# 6 Conclusions

Wind tunnel measurements of non-rotating porous disk models have demonstrated that yawing produces a wake deflection consistent with what is expected for rotating real wind turbines. Furthermore, we have observed the formation of a curled wake, a phenomenon which has not yet been described in previous studies of wind turbines in yaw. The curling of the wake is consistent with the distributions of spanwise mean velocity in the wake, which shifts the wake defect velocity more strongly sideways near the wake center than at the top and bottom, where it is shifted in the opposite direction. Asymmetries and wake deformations have been previously described as a result of Ekman layer transverse shear in the atmospheric boundary layer [47, 48].

LES results using both actuator line and actuator disk wind turbine models confirm the experimental observations qualitatively. Quantitatively, significant differences exist because the simulated wakes under uniform laminar inflow do not transition quickly to a turbulent state (the resolution used was too coarse to resolve individual bars in the grid from which the disks were made). However, both simulations and experiments, are able to confirm the existence of a *curled wake* phenomena for a yawed wind turbine under uniform inflow. The illustration in Fig. 13 summarizes the curled wake morphology as observed in our results. As the wake evolves donwstream, a set of counter rotating vortices created by the yawed turbine, deform the wake, giving it its *curled wake* shape.

The curled wake shape has potential implications for the power optimization of wind farms using yawed wake deflection. Importantly, some previous studies have only considered XY planes at hub height to characterize the deflection of a turbine wake for the purposes of optimization. However, present data show that the wake of a yawed turbine exhibits asymmetry in 3D, and that such 3D effects must be considered to better understand the effects of yaw as a wake deflection tool. Specifically, the curling may cause a wake to miss more of a downstream turbine as implied only by the deflection as measured by $y_c(x)$, since it may "wrap" around the downstream rotor [49].

Future experiments should study the decay of curled wakes under turbulent inflow conditions, more relevant to atmospheric boundary layer conditions. It is possible that the turbulent diffusion of the wake curling depends on the turbulence intensity and thus the latter may be an important parameter for control also when attempting to include the wake curling phenomenon in power predictions. It also remains to study and verify the wake curling phenomenon in field studies.

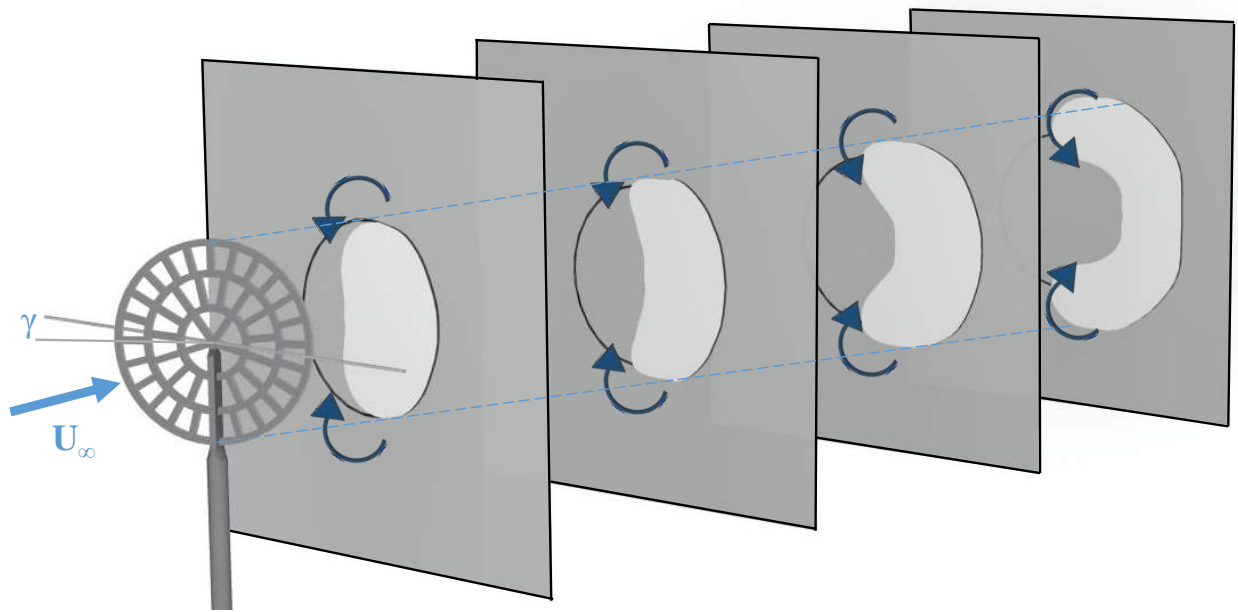

Figure 13: Wake Shape Deformation Sketch. Yaw angle is denoted as $\gamma$, shown as 30°. The deformed wake shapes are shown on dark grey successive downstream planes in light grey, deformed by the counter rotating vortex pair. The black circles show the turbine rotor area on each plane.

*Acknowledgements:* This work has been funded by the National Science Foundation (grants CBET-113380 and IIA-1243482, the WINDINSPIRE project). JB and JM are supported by ERC (ActiveWind-Farms, grant no. 306471).

# Appendix

The formulation based on Ref. [21] for the wake deflection of a turbine in yaw is as follows:

$$\alpha(x) = \frac{dy_c}{dx} = \frac{cos^2(\gamma)sin(\gamma)C_T/2}{(1 + \beta \cdot x/D)^2}, \quad (3)$$

where $\alpha$ is the wake skew angle, $\beta$ is the wake expansion factor (we used $\beta = 2k = 0.1$ since k = 0.03 - 0.06 has been shown to be a representative range [50]. Integrating in x, as also done in Ref. [25] and using $y_c(x = 0) = 0$ leads to

$$\frac{y_c(x)}{D} = cos^2(\gamma)sin(\gamma)\frac{C_T}{2}\frac{1}{\beta} \cdot \left(1 - \frac{1}{\beta \cdot x/D + 1}\right). \quad (4)$$

# References

[1] W. Short and N. Blair, "20% wind energy by 2030: Increasing wind energy's contribution to US electricity supply," *DOE Report No. DOfJGO-102008-2567*, 2008.

[2] G. W. E. Council, "Global wind energy outlook 2012," *GWEC, November*, 2012.

[3] S. Frandsen, "On the wind speed reduction in the center of large clusters of wind turbines," *J. of Wind Engineering & industrial aerodynamics*, vol. 39, pp. 251–265, 1992.

[4] S. Frandsen, R. Barthelmie, S. Pryor, O. Rathmann, S. Larsen, J. Højstrup, and M. Thøgersen, "Analytical modelling of wind speed deficit in large offshore wind farms," *Wind energy*, vol. 9, pp. 39–53, 2006.

[5] R. J. Barthelmie and L. E. Jensen, "Evaluation of wind farm efficiency and wind turbine wakes at the Nysted offshore wind farm," *Wind Energy*, vol. 13, no. 6, pp. 573–586, 2010.

[6] K. S. Hansen, R. J. Barthelmie, L. E. Jensen, and A. Sommer, "The impact of turbulence intensity and atmospheric stability on power deficits due to wind turbine wakes at horns rev wind farm," *Wind Energy*, vol. 15, pp. 183–196, 2012.

[7] T. Ackermann, *Wind power in power systems.* John Wiley & Sons, 2005.

[8] M. Abdullah, A. Yatim, C. Tan, and R. Saidur, "A review of maximum power point tracking algorithms for wind energy systems," *Renewable and Sustainable Energy Reviews*, vol. 16, no. 5, pp. 3220–3227, 2012.

[9] P. A. Fleming, P. M. Gebraad, S. Lee, J.-W. van Wingerden, K. Johnson, M. Churchfield, J. Michalakes, P. Spalart, and P. Moriarty, "Evaluating techniques for redirecting turbine wakes using SOWFA," *Renewable Energy*, vol. 70, pp. 211–218, 2014.

[10] P. A. Fleming, A. Ning, P. M. Gebraad, and K. Dykes, "Wind plant system engineering through optimization of layout and yaw control," *Wind Energy*, 2015.

[11] J. P. Goit and J. Meyers, "Optimal control of energy extraction in wind-farm boundary layers," *Journal of Fluid Mechanics*, vol. 768, pp. 5–50, 2015.

[12] I. Grant, P. Parkin, and X. Wang, "Optical vortex tracking studies of a horizontal axis wind turbine in yaw using laser-sheet, flow visualisation," *Experiments in Fluids*, vol. 23, no. 6, pp. 513–519, 1997.

[13] P. Parkin, R. Holm, and D. Medici, "The application of PIV to the wake of a wind turbine in yaw," in *Particle Image Velocimetry; Gottingen; 17 September 2001 through 19 September 2001*, pp. 155–162, 2001.

[14] M. Adaramola and P.-Å. Krogstad, "Experimental investigation of wake effects on wind turbine performance," *Renewable Energy*, vol. 36, no. 8, pp. 2078–2086, 2011.

[15] P. Fleming, P. Gebraad, J.-W. van Wingerden, S. Lee, M. Churchfield, A. Scholbrock, J. Michalakes, K. Johnson, and P. Moriarty, "The SOWFA super-controller: A high-fidelity tool for evaluating wind plant control approaches," in *Proceedings of the EWEA Annual Meeting, Vienna, Austria*, 2013.

[16] J. Park, S. Kwon, and K. H. Law, "Wind farm power maximization based on a cooperative static game approach," in *SPIE Smart Structures and Materials+ Nondestructive Evaluation and Health Monitoring*, pp. 86880R–86880R, International Society for Optics and Photonics, 2013.

[17] T. Mikkelsen, N. Angelou, K. Hansen, M. Sjöholm, M. Harris, C. Slinger, P. Hadley, R. Scullion, G. Ellis, and G. Vives, "A spinner-integrated wind lidar for enhanced wind turbine control," *Wind Energy*, vol. 16, no. 4, pp. 625–643, 2013.

[18] J. Wagenaar, L. Machielse, and J. Schepers, "Controlling wind in ECN's scaled wind farm," *Proc. Europe Premier Wind Energy Event*, pp. 685–694, 2012.

[19] K. A. Kragh and M. H. Hansen, "Load alleviation of wind turbines by yaw misalignment," *Wind Energy*, vol. 17, no. 7, pp. 971–982, 2014.

[20] M. Bastankhah and F. Porté-Agel, "A wind-tunnel investigation of wind-turbine wakes in yawed conditions," in *Journal of Physics: Conference Series*, vol. 625, p. 012014, IOP Publishing, 2015.

[21] A. Jiménez, A. Crespo, and E. Migoya, "Application of a LES technique to characterize the wake deflection of a wind turbine in yaw," *Wind Energy*, vol. 13, pp. 559–572, 2010.

[22] C. Tsalicoglou, S. Jafari, N. Chokani, and R. S. Abhari, "RANS Computations of MEXICO Rotor in Uniform and Yawed Inflow," *Journal of Engineering for Gas Turbines and Power*, vol. 136, no. 1, p. 011202, 2014.

[23] S. Guntur, N. Troldborg, and M. Gaunaa, "Application of engineering models to predict wake deflection due to a tilted wind turbine," in *EWEA 2012-European Wind Energy Conference & Exhibition*, 2012.

[24] J. M. Jonkman, S. Butterfield, W. Musial, and G. Scott, *Definition of a 5-MW reference wind turbine for offshore system development*. National Renewable Energy Laboratory Golden, CO, 2009.

[25] L. Luo, N. Srivastava, and P. Ramaprabhu, "A study of intensified wake deflection by multiple yawed turbines based on large eddy simulations," 2014.

[26] D. Medici and P. Alfredsson, "Measurements on a wind turbine wake: 3D effects and bluff body vortex shedding," *Wind Energy*, vol. 9, no. 3, pp. 219–236, 2006.

[27] P. Fleming, A. Scholbrock, A. Jehu, S. Davoust, E. Osler, A. Wright, and A. Clifton, "Field-test results using a nacelle-mounted lidar for improving wind turbine power capture by reducing yaw misalignment," in *Journal of Physics: Conference Series*, vol. 524, p. 012002, IOP Publishing, 2014.

[28] J. Bossuyt, M. Howland, C. Meneveau, and J. Meyers, "Measuring power output intermittency and unsteady loading in a micro wind farm model," in *34th Wind Energy Symposium*, pp. http://dx.doi.org/10.2514/6.2016–1992, 2016.

[29] A. Thormann and C. Meneveau, "Decaying turbulence in the presence of a shearless uniform kinetic energy gradient," *Journal of Turbulence*, vol. 16, no. 5, pp. 442–459, 2015.

[30] K. Talluru, V. Kulandaivelu, N. Hutchins, and I. Marusic, "A calibration technique to correct sensor drift issues in hot-wire anemometry," *Measurement Science and Technology*, vol. 25, no. 10, p. 105304, 2014.

[31] S. Dhawan and R. Narasimha, "Some properties of boundary layer flow during the transition from laminar to turbulent motion," *Journal of Fluid Mechanics*, vol. 3, no. 04, pp. 418–436, 1958.

[32] N. Hutchins, T. B. Nickels, I. Marusic, and M. Chong, "Hot-wire spatial resolution issues in wall-bounded turbulence," *Journal of Fluid Mechanics*, vol. 635, pp. 103–136, 2009.

[33] J.-J. Trujillo, F. Bingöl, G. C. Larsen, J. Mann, and M. Kühn, "Light detection and ranging measurements of wake dynamics. Part II: two-dimensional scanning," *Wind Energy*, vol. 14, no. 1, pp. 61–75, 2011.

[34] P. Fleming, P. Gebraad, M. Churchfield, J. van Wingerden, A. Scholbrock, and P. Moriarty, "Using particle filters to track wind turbine wakes for improved wind plant controls," in *American Control Conference (ACC), 2014*, pp. 3734–3741, IEEE, 2014.

[35] M. Calaf, C. Meneveau, and J. Meyers, "Large eddy simulations of fully developed wind-turbine array boundary layers," *Phys. Fluids*, vol. 22, p. 015110, 2010.

[36] M. Calaf, M. B. Parlange, and C. Meneveau, "Large eddy simulation study of scalar transport in fully developed wind-turbine array boundary layers," *Phys. Fluids*, vol. 23, p. 126603, 2011.

[37] R. J. A. M. Stevens, J. Graham, and C. Meneveau, "A concurrent precursor inflow method for Large Eddy Simulations and applications to finite length wind farms," *Renewable Energy*, vol. 68, pp. 46–50, 2014.

[38] R. Stevens, D. Gayme, and C. Meneveau, "Large eddy simulation studies of the effects of alignment and wind farm length," *J. Renewable and Sustainable Energy*, vol. 6, no. 2, p. 023105, 2014.

[39] L. Martínez-Tossas, M. Churchfield, and C. Meneveau, "Large eddy simulation of wind turbine wakes: detailed comparisons of two codes focusing on effects of numerics and subgrid modeling," in *Proc. EAWEA Wake Conf. June 2015, Visby, Sweden*.

[40] E. Bou-Zeid, C. Meneveau, and M. B. Parlange, "A scale-dependent Lagrangian dynamic model for large eddy simulation of complex turbulent flows," *Phys. Fluids*, vol. 17, p. 025105, 2005.

[41] H. Sarlak, C. Meneveau, and J. N. Sørensen, "Role of subgrid-scale modeling in large eddy simulation of wind turbine wake interactions," *Renewable Energy*, vol. 77, pp. 386–399, 2015.

[42] L. A. M.-T. Tossas, R. J. Stevens, and C. Meneveau, "Wind Turbine Large-Eddy Simulations on Very Coarse Grid Resolutions using an Actuator Line Model," pp. http://dx.doi.org/10.2514/6.2016–1261, 2016.

[43] C. VerHulst and C. Meneveau, "Large eddy simulation study of the kinetic energy entrainment by energetic turbulent flow structures in large wind farms," *Phys. Fluids*, vol. 26, no. 2, p. 025113, 2014.

[44] B. R. Munson, D. F. Young, and T. H. Okiishi, *Fundamentals of fluid mechanics*. New York, 1990.

[45] J. N. Sørensen and W. Z. Shen, "Numerical modeling of wind turbine wakes," *Journal of fluids engineering*, vol. 124, no. 2, pp. 393–399, 2002.

[46] L. A. Martínez-Tossas, M. J. Churchfield, and S. Leonardi, "Large eddy simulations of the flow past wind turbines: actuator line and disk modeling," *Wind Energy*, vol. 18, no. 6, pp. 1047–1060, 2015.

[47] N. Zhou, J. Chen, D. E. Adams, and S. Fleeter, "Influence of inflow conditions on turbine loading and wake structures predicted by large eddy simulations using exact geometry," *Wind Energy*, 2015.

[48] N. Sezer-Uzol and O. Uzol, "Effect of steady and transient wind shear on the wake structure and performance of a horizontal axis wind turbine rotor," *Wind Energy*, vol. 16, no. 1, pp. 1–17, 2013.

[49] L. A. Martínez-Tossas, M. Howland, and C. Meneveau, "Large eddy simulation of wind turbine wakes with yaw effects," *APS DFD GFM 2015*, p. DOI: http://dx.doi.org/10.1103/APS.DFD.2015.GFM.V0012.

[50] R. J. Stevens, D. F. Gayme, and C. Meneveau, "Coupled wake boundary layer model of wind-farms," *Journal of Renewable and Sustainable Energy*, vol. 7, no. 2, p. 023115, 2015.